\documentclass{JINST}

\title{MegaPipe astrometry for the New Horizons spacecraft}

\author{Stephen D. J. Gwyn$^a$\thanks{Corresponding author.}\\
\llap{$^a$}Canadian Astronomy Data Centre,\\
5071 West Saanich Rd, Victoria BC, V9E 2E7, Canada\\  
E-mail: \email{stephen.gwyn@nrc.ca}}

\abstract{
The New Horizons spacecraft, launched by NASA in 2006, will arrive in
the Pluto-Charon system on July 14, 2015. There, it will spend a few
hours imaging Pluto and its moons. It will then have a small amount of
reserve propellant which will be used to direct the probe on to a
second, yet to be discovered object in the Kuiper Belt. Data from the
MegaPrime camera on CFHT was used to build a precise, high density
astrometric reference frame for both the final approach into the Pluto
system and the search for the secondary target. Pluto currently lies
in the galactic plane. This is a hindrance in that there are potential
problems with confusion. However, it also a benefit, since it allows the
use of the UCAC4 astrometric reference catalog, which is normally too
sparse for use with MegaCam images.  The astrometric accuracy of the
final catalogs, as measured by the residuals, is 0.02 arcseconds.}


\begin{document}

\section{Introduction}\label{sec:intro}

MegaCam \cite{megacam} is a mosaic camera with a one degree field of
view on the Canada France Hawaii Telescope (CFHT). It consists of 36
E2V CCD detectors.  MegaPipe \cite{megapipe} is the MegaCam data
processing pipeline at the Canadian Astronomy Data Center. MegaPipe
began processing all publicly available MegaCam data in 2005. Since
then, it has produced over 3000 square degrees of stacked images and
catalogs. It also processes PI data on request and handles data from
the CFHT Large Programs.

The New Horizons \cite{newhorizons} spacecraft was launched by NASA in
2006 and is currently en route to Pluto.  In 2015, on July 14, it will
fly through the Pluto-Charon system for a few hours.  After the Pluto
flyby, the space craft will have a small amount of reserve
propellant. This will be used to alter course toward a second Kuiper Belt
Object (KBO). The ongoing search \cite{seckbo} for this object involves data
from the Subaru and Magellan as well as CFHT. The MegaPipe images are
being used as the astrometric reference frame for the search. Between
45 and 6 days before the encounter, a search will be conducted of the
Pluto-Charon system for potentially hazardous debris using LORRI, the
onboard detector. Again, MegaPipe data is being used as the
astrometric reference for this search.

\begin{figure}[tbp] 
\centering
\includegraphics[width=.5\textwidth]{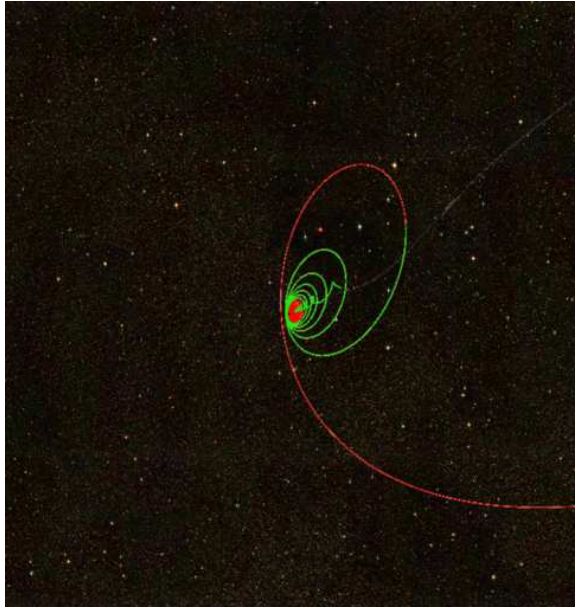}
\caption{This figure illustrates the approach of New Horizons into the Pluto system.
The line shows the position of Pluto, as seen by the space craft, with
respect to the background stars. The colors indicate times: green
indicates 45 and 6 days pre-encounter when the search for hazardous
debris will occur. The image is one degree on a side.}
\label{fig:approach}
\end{figure}

This paper describes the production of the astrometric reference catalogs used
for both the search for the second KBO target and the debris search.  The
construction of these catalogs was complicated by the fact that Pluto
is currently crossing the galactic plane. The stellar density in this
region of the sky is extremely high for the most part, but drops off
sharply in places due to the presence of dust clouds in the field, as
shown in figure \ref{fig:approach}. There is a strong potential for
confusion when matching stars in the images to stars in the astrometric reference
catalog.

\section{Method}

The MegaPipe astrometric calibration method has been described
elsewhere (e.g., \cite{megapipe}, \cite{fraseroccult}). Consequently,
the following discussion will be brief, except to emphasize the
inherent strengths of the standard MegaPipe procedures when dealing
with crowded images and what recent improvements have been
incorporated to those procedures to deal with the crowded New Horizons
field.

The input images from MegaCam have been detrended with Elixir
\cite{elixir} and contain an initial WCS (World Coordinate System)
which is used as a starting point. An obverved source catalog is
generated for each image with SExtractor\cite{sex}. MegaCam images are
typically much deeper than the external astrometric references
catalogs; a magnitude cut is applied to the observed source catalog to
minimize the number of potentially confusing sources.

The observed source catalog is matched to the external astrometric
reference catalog under the assumption the Elixir linear WCS is close
to correct in terms orientation and scaling.  Small shifts in RA and
Dec are applied to the WCS until the number of matches between the
observed source catalog and the external astrometric reference catalog
is maximized. When the geometry of the detectors is well known, this
simple approach is sufficient, and is more robust in crowded fields
than the more sophisticated Fourier transform methods or triangle- and
quad-matching methods.

The number of catalog matches is rough indicator of the quality of
the WCS match. Normally, off the galactic plane, for a single MegaCam CCD, 30
matches between the external astrometric reference catalog and the observed catalog
indicates a successful identification. In the crowded New Horizons
fields, the minimum match number was set to 300 matches per CCD.  CCDs
with fewer matches were flagged for visual inspection. Most such cases
the CCD was found to lie on a patch of higher obscuration (i.e. fewer stars).
The few remaining cases showed spurious matches and had to be
rematched.

\begin{figure}[tbp] 
\label{fig:ucac2mass}
\includegraphics[width=1\textwidth]{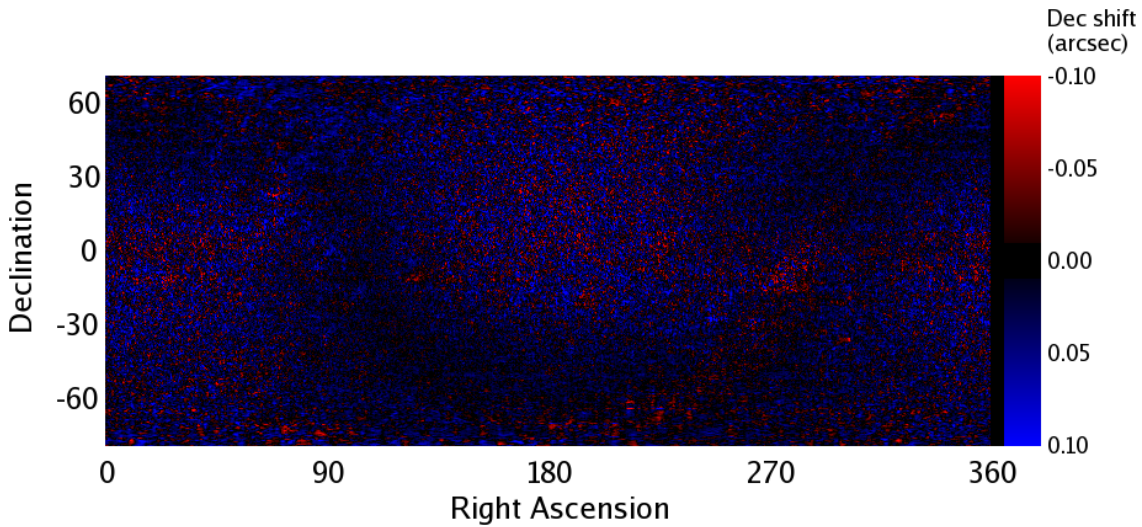}
\caption{The astrometric residuals between 2MASS and UCAC4 in
Declination.  The color coding indicates the size of the astrometric
residuals. The galactic plane is visible as a darker swath
inidicating smaller astrometric residuals. More subtly, there is a
pattern with a 6 degree period in declination, indicating problems
with the 2MASS catalog.}
\label{fig:ucac42mass}
\end{figure}

Normally, MegaPipe uses the 2MASS \cite{2mass} catalog as the external
astrometric reference, because its high source density. However,
since the New Horizons fields are near the galactic center, a
shallower but higher precision external astrometric reference catalog can used, namely
UCAC4 \cite{ucac4}.  When the 2MASS and UCAC4 catalogs are compared
over the whole sky, small systematic shifts are visible, as shown in
figure \ref{fig:ucac2mass}.  The pattern of the offsets have a 6
degree periodicity in Declination, matching the 2MASS observation
strategy, indicating that there may be errors in the 2MASS
catalog. Similar, but smaller offsets are visible when the SDSS
DR9 \cite{sdssdr9} and UCAC4 catalogs are compared, with the pattern
matching the SDSS swathes.

Once the observed source catalog and the external astrometric reference catalog are
matched, the field distortion can be measured. MegaPipe models the
focal plane distortion as
\begin{equation}
\label{eqn:distort}
R=r (1+a_2r^2+a_4r^4)
\end{equation}
where $r$ is the measured radius from the center and $R$ is the true
radius. This means that only two coefficients ($a_2$ and $a_4$), which
can be determined once for the whole mosaic, are needed to describe
the distortion. In typical astrometric packages, a third-order
polynomial in $x$ and $y$ pixel position is computed, requiring 20
parameters for each detector of a mosaic. In the case of MegaCam, this
would mean 720 coefficients. MegaPipe instead only models the linear
part of the WCS at the chip level. This means that there only 6
parameters per chip, plus 2 for the higher order distortion, for a
total of 218. Keeping the number of parameters relatively small
prevents spurious matches in crowded fields, and allows more robust
fitting when there are steep gradients in the source density of the
external astrometric reference catalog. The values of $a_2$ and $a_4$
are calculated for each MegaCam exposure, although the values change
only very slightly from between exposures. The values of $a_2$ and
$a_4$ are initially set to nominal values and the linear terms for
each chip are computed by linear regression. The astrometric residuals
are computed using the current fit. The values of $a_2$ and $a_4$ are
then allowed to vary until the astrometric residuals are minimized.

While the distortion is expressed internally in MegaPipe as shown in
equation \ref{eqn:distort}, it is expressed as a polynomial using the
PV\underline{~}nn WCS keywords in the FITS headers.  This conversion
merely re-expresses the distortion; no additional fitting is actually
done, despite the apparent increase in the number of coefficients.

This initial matching and fitting procedure is applied to the input
images.  After the WCS has been computed, it is applied to the
observed source catalogs to convert the $x,y$ pixel coordinates to RA and
Dec. The RA/Dec catalogs are then combined to produce an intermediate
astrometric catalog covering the whole field. Sources common to more
than one catalog are identified and their coordinates from different
images averaged to increase the positional accuracy.

The matching and fitting procedure is repeated on the input images,
but this time, instead of using UCAC4 as the external astrometric reference,
the intermediate catalog is used. The images, with updated WCS in
their headers are combined using
SWarp\footnote{http://www.astromatic.net/software/swarp}.  SExtractor
is run on the combined stack to generate the final astrometric
catalog. The SExtractor parameters are changed slightly relative to the
default MegaPipe configuration to deal with the crowding.  In
particular, the minimum contrast parameter for deblending, {\tt
DEBLEND\underline{~}MINCONT}, is set to a much lower value.  This image
combination step effectively merges the astrometric measurements from
each image at the pixel level. This final astrometric catalog
can be used to calibrate the individual input images a final time
so that they can be used for KBO searches.

\begin{figure}[tbp] 
\centering
\includegraphics[width=.5\textwidth]{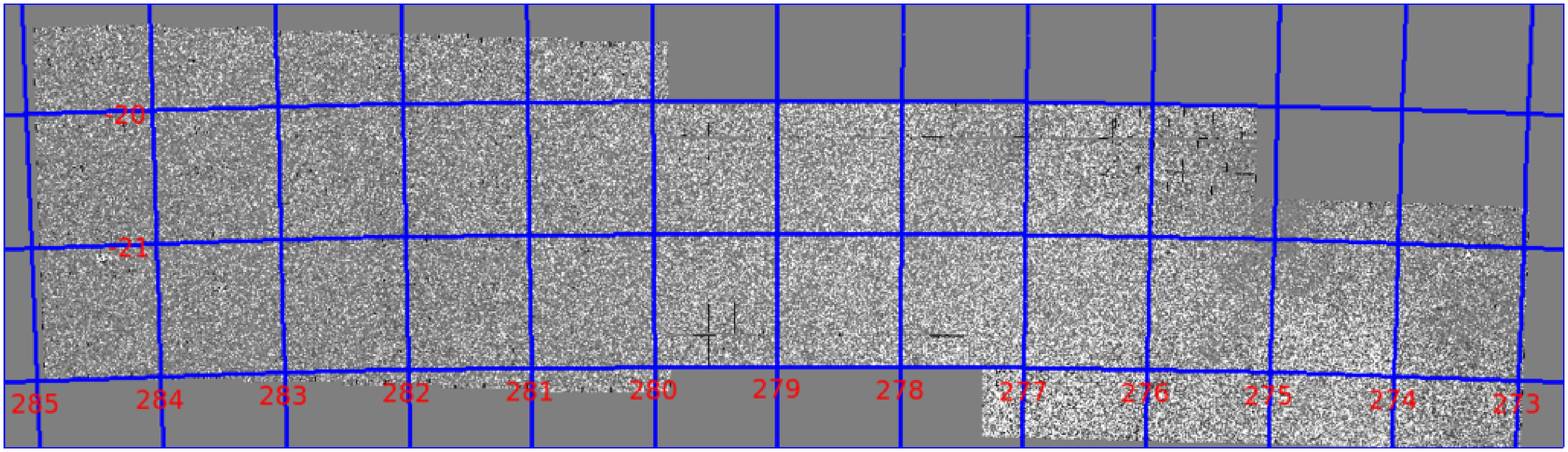}
\caption{This image shows the 20 square degrees that are being searched for a secondary
KBO target for the New Horizons probe.}
\label{fig:kbosearch}
\end{figure}

This method was applied to the KBO search fields. The resulting image
is shown in figure \ref{fig:kbosearch}.

The Pluto hazard detection field has data taken in multiple bands.
The images taken in the r-band were processed first. The $g$- and $i$-
band data were calibrated using the r-band image as the astrometric
reference. Despite the difference in wavelength between the three
bands, chromatic differential refraction will not introduce any
systematic effects in the astrometry of the $g$ and $i$ images
relative to the $r$-band image as a function of position. The
positions of stars will shift in azimuthal direction function of their
color, but because the images were taken a low airmass, the amount is
small, $\sim0.004''$ per magnitude in $(g-i)$.  The gri images were
combined to produce the color image indicating the background in
figure \ref{fig:approach}.

\section{Tests}

\begin{figure}[tbp] 
\centering
\includegraphics[width=0.7\textwidth]{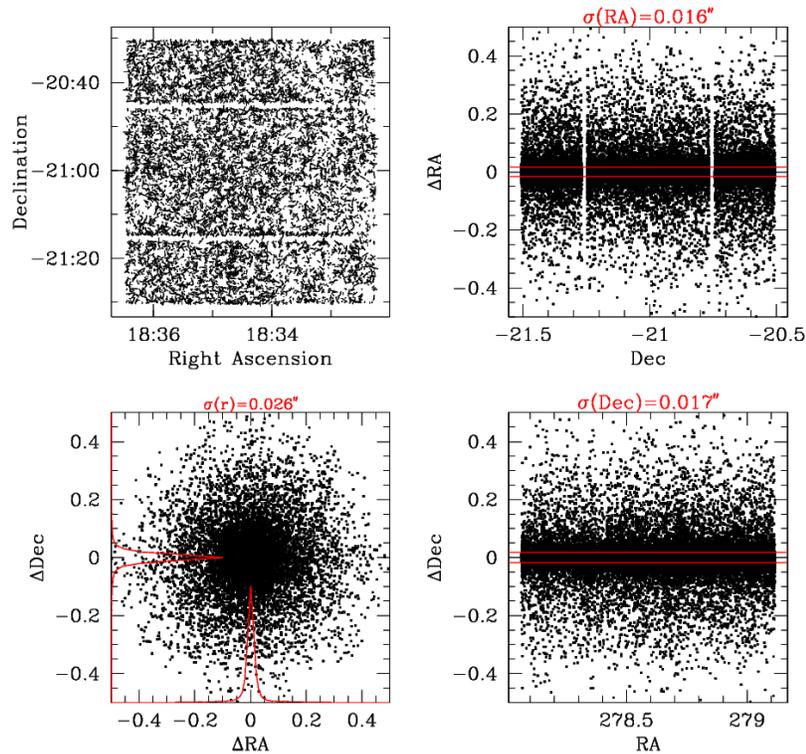}
\caption{The top left panel shows the astrometric residuals as a whisker plot.
The lines show the orientation and relative amplitude of the
residuals.  The amplitude of the residuals is greatly exaggerated. The
bottom left panel shows the residuals in RA and Dec as a scatter
plot. The panels on the right show the residuals in RA and Dec as
function of Dec and RA respectively. While the random errors are on
the order of 0.02$''$, there is no evidence for any systematic errors.
}
\label{fig:astresex}
\end{figure}

The catalogs from the calibrated individual images were matched
against each other and against the final catalogs and the residuals
examined. For the Pluto encounter field, the catalogs in the gri bands
were also checked against each other. The astrometric residuals were
typically 0.02$''$, as shown in figure \ref{fig:astresex}.  The
unmodified MegaPipe pipeline produces astrometric residuals that are
typically 0.04$''$. Systematic errors would be readily detectable in
this figure, particularly in the whisker plot in the upper left panel,
where they would show up as groups of parallel lines.

\begin{figure}[tbp] 
\centering
\includegraphics[width=0.75\textwidth]{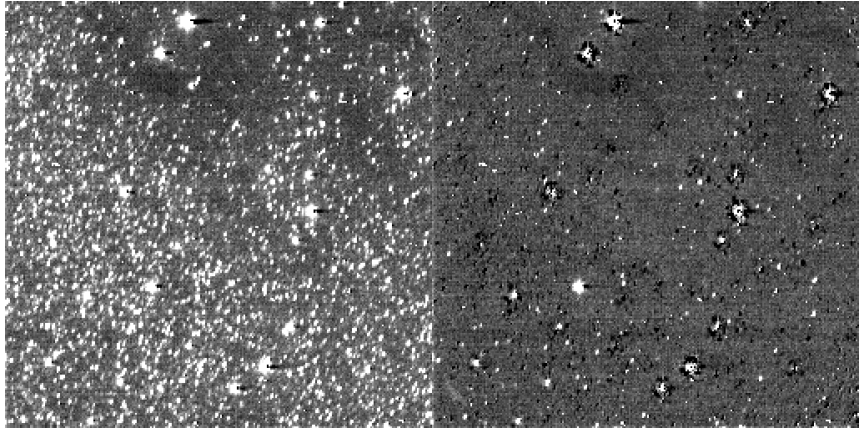}
\caption{The left panel shows an image taken with LORRI of the Pluto
  approach field.  The right panel shows the residual between this
  image and a simulated image constructed from by placing predicted
  LORRI PSFs at the positions and flux levels of objects in the
  MegaCam astrometric (and photometric) catalog. Most of the stars
  completely disappear, indicating good astrometry. The remaining
  residuals are centered on the stars, and indicate that there is a
  mismatch between the MegaCam and LORRI passbands.}
\label{fig:catmatch}
\end{figure}

The positions and magnitudes of the Pluto encounter field catalog were
used to simulate an image taken with LORRI.  The catalog positions
were convolved with LORRI's PSF.  The simulated image was subtracted
from a real image, after making small adjustments for the spacecraft's
pointing and roll angle. Although small residuals (both positive and
negative) reveal a bandpass mismatch between MegaCam and LORRI, the
alignment is perfect, as shown in figure \ref{fig:catmatch}.

\bibliography{stan}

\end{document}